\documentclass[12pt]{article}
\usepackage{amssymb}
\usepackage{amsfonts}
\usepackage{amsmath}

\input{tcilatex}

\begin{document}

\title{BRST superspace and auxiliary fields for $\mathcal{N=}$1
supersymmetric Yang-Mills theory}
\author{A. Aidaoui, A. Meziane\thanks{%
E-mail: meziane@univ-oran.dz} \ and \ M. Tahiri \\
Laboratoire de Physique Th\'{e}orique\\
Universit\'{e} d'Oran Es-senia, 31100 Oran, \ Algeria\\
}
\maketitle

\begin{abstract}
We use a Becchi-Rouet-Stora-Tyutin (BRST) superspace approach \ to formulate
off-shell nilpotent BRST and anti-BRST transformations in four dimensional $%
\mathcal{N=}$1 supersymmetric Yang-Mills theory. The method is based on the
possibility of introducing auxiliary fields through the supersymmetric
transformations of the superpartener of the gauge potential associated to a
supersymmetric Yang-Mills connection. These fields are required to achieve
the off-shell nilpotency of the BRST and anti-BRST operators. We also show \
how this off-shell structure is used to build the BRST and anti--BRST
invariant gauge-fixing \ quantum action.

BRST and anti-BRST algebra.

Pacs No.: 11.15.-q
\end{abstract}

\section{ Introduction}

An on-shell quantization of general gauge theories which are reductible
and/or whose classical gauge algebra is not closed (for a review see e.g.
Ref. \cite{Henneaux}), can be successfully performed by the
Batalin-Vilkovisky (BV) formalism \cite{batalin22}. The BV formalism is a
very general covariant Lagrangian approach which overcomes the need of
closed classical algebra by a suitable construction of BRST operator. The
construction is realized by introducing a set of new antifields besides the
fields occurring in the theory. The elimination of these antifields at the
quantum level via a gauge-fixing procedure leads to the quantum theory in
which effective BRST transformations\ are nilpotent only on-shell.

Another possibility for quantizing reductible and/or open gauge theories
consists to introduce a set of auxiliary fields\ as in supersymmetric
theories, such as the Wess-Zumino model for which it is only with auxiliary
fields that one can obtain a tensor calculus \cite{Van-Nieuwen}, or in
topological antisymmetric tensor gauge theories, the so-called BF theories 
\cite{tahiri2}. The introduction of auxiliary fields are needed to close the
gauge algebra, and then provides an efficient way to use the usual
Faddeev-Popov quantization method \cite{becchi}.

On the other hand, it is well known that in the superspace formalism one
can\ naturally introduce a set of auxiliary fields that gives rise to the
construction of the off-shell BRST invariant quantum action. Indeed, as
shown in Ref. \cite{tahiri2} for the case of $4D$ non-Abelian BF theory and
Ref. \cite{tahiri22} for the case of the simple supergravity where the
classical gauge algebra is open \cite{Van-Nieuwen}, the superspace formalism
has been used in order to realize the quantization of such theories. It
leads to introduce the minimal set of auxiliary fields ensuring the
off-shell invariance of the quantum action.

In the context of superspace formalism, the gauge field, the ghost and
anti-ghost fields in gauge theories can be incorporated into a natural gauge
superconnection by extending spacetime to a $(4,2)-$dimensional superspace 
\cite{Bonora}. In this framework, the BRST and anti-BRST transformations can
be derived by imposing horizontality conditions on the supercurvature
associated to a superconnection on a superspace.

Let us mention that in the case of supersymmetric theories the corresponding
supersymmetric algebra close only modulo equations of motion \cite%
{Sohnius,Wess}. So, the appropriate framework to quantize such theories is
the BV\ one. Indeed, in Ref. \cite{Baulieu}\ it has been discussed how to
realize the quantization of supersymmetric systems in BV approach.

As shown in \cite{Lavrov1,Lavrov2}, a superfield description of the BV
quantization method can be realized by simply introducing superfields whose
lowest components coincide with the usual fields in the BV formalism.
Superfield method has provided a powerful tool for producing supersymmetric
field equations for any degree of supersymmetry. In \cite{Sohnius,WEinberg},
one has also established an off-shell superfield formulation of four
dimensional $\mathcal{N}=1$ supersymmetric Yang-Mills theory. Considerably
more involved off-shell superfield formulations are also available for $%
\mathcal{N}=2$ in terms of harmonic and analytic superspace \cite{Galperin},
while the off-shell formulation of $\mathcal{N}=4$ supersymmetric Yang-Mills
theory with non-Abelian gauge group $SU(N)$ is not available in terms of
unconstrained fields because it would require the introduction of an
infinite set of auxiliary fields \cite{Falk}.

The purpose of the present paper is to derive, in the framework of BRST
superspace\footnote{%
Here, we call the superspace obtained by enlarging spacetime with two
ordinary anticommuting coordinates BRST superspace, in order to distinguish
it from superspace of supersymmetric theories.}, the off-shell nilpotent
version of the BRST and anti-BRST transformations for $\mathcal{N}=1,$\ $D=4$
supersymmetric Yang-Mills theory $(SYM_{4})$, in analogy to what is realized
for the case of\ simple supergravity \cite{tahiri2} and the four-dimensional
non-Abelian BF theory \cite{tahiri22}. The Action for the $\mathcal{N}=1$\
supersymmetric Yang-Mills in four dimensions is given by

\begin{equation}
S_{0}=\int dx^{4}\left( \frac{-1}{2g^{2}}trF_{\mu \nu }F^{\mu \nu }+\frac{%
\theta }{8\pi ^{2}}trF_{\mu \nu }\widetilde{F}^{\mu \nu }-\frac{i}{2}tr%
\overline{\lambda }\sigma ^{\mu }D_{\mu }\lambda \right) ,
\label{Lagrangian}
\end{equation}%
where $(A_{\mu },\lambda )$ is the gauge multiplet, $g$ is the gauge
coupling, $\theta $ is the instanton angle, the field strength is $F_{\mu
\nu }=\partial _{\mu }A_{\nu }-\partial _{\nu }A_{\mu }+\left[ A_{\mu
},A_{\nu }\right] ,$ $\widetilde{F}_{\mu \nu }=\frac{1}{2}\epsilon _{\mu \nu
\rho \sigma }F^{\rho \sigma }$ is the dual of $F,$ and $D_{\mu }=\partial
_{\mu }+\left[ A_{\mu },\right] $. By construction, the Action (\ref%
{Lagrangian}) is invariant under the supersymmetry transformations

\begin{eqnarray}
\delta _{\xi }A_{\mu } &=&i\overline{\xi }\overline{\sigma }_{\mu }\lambda -i%
\overline{\lambda }\overline{\sigma }_{\mu }\xi ,  \notag \\
\delta _{\xi }\lambda &=&\sigma ^{\mu \nu }F_{\mu \nu }\xi ,
\label{Supersy transf}
\end{eqnarray}%
where $\xi $ is a spin $1/2$ valued infinitesimal supersymmetry parameter.

Our paper is organized as follows. In Section 2, the BRST superspace
approach and horizontality conditions for $\mathcal{N}=1,$\ $D=4$
supersymmetric Yang-Mills theory are discussed. In Section 3, we introduce\
the various fields and derive the off-shell nilpotent\ BRST and anti-BRST
transformations. The construction\ of the BRST-invariant quantum action for $%
\mathcal{N}=1$ super Yang-Mills theory in terms of this off-shell structure
is described in Section 4. The obtained quantum action permits us to see
that the extrafields are nondynamical, auxiliary fields. Section 5 is
devoted to concluding remarks.

\section{BRST superspace}

Let $\Phi $\ be a super Yang-Mills connection on a $(4,2)-$dimensional BRST
superspace with coordinates $z=(z^{M})=(x^{\mu },\theta ^{\alpha }),$\ where 
$(x^{\mu })_{\mu =1,..,4}$\ are the coordinates of the spacetime manifolds
and $(\theta ^{\alpha })_{\alpha =1,2}$\ are ordinary anticommuting
variables. Acting the exterior covariant superdifferential $D$ on $\Phi $ we
obtain the supercurvature $\Omega $ satisfying the structure equation, $%
\Omega =d\Phi +\frac{1}{2}\left[ \Phi ,\Phi \right] ,$ and the Bianchi
identity, $d\Omega +\left[ \Phi ,\Omega \right] =0.$ The superconnection $%
\Phi $\ as 1-superform on the BRST\ superspace can be written as

\begin{equation}
\Phi =dz^{M}(\Phi _{M}^{i}I_{i}+\Phi _{M}^{\mu }P_{\mu }+\Phi _{M}^{\alpha
}Q_{\alpha }),  \label{eq1a}
\end{equation}%
where $\{I_{i}\}_{i=1,...,d},$ and $\{P_{\mu },Q_{\alpha }\}_{\mu
=1,...,4;\alpha =1,...,4}$ are the generators of the internal symmetry group 
$(G)$ and the $\mathcal{N=}1$ supersymmetric group $(SG)$ respectively. They
satisfy the following commutation relations

\begin{eqnarray}
\left[ I_{i},I_{j}\right] &=&f_{ij}^{k}I_{k},  \notag \\
\left[ I_{i},P_{\mu }\right] &=&\left[ Q_{\alpha },P_{\mu }\right] =\left[
P_{\nu },P_{\mu }\right] =0,  \notag \\
\left[ Q_{\alpha },Q_{\beta }\right] &=&2(\gamma ^{\mu })_{\alpha \beta
}P_{\mu },  \notag \\
\left[ I_{i},Q_{\alpha }\right] &=&b_{i}^{\ast }Q_{\alpha },  \label{eq2}
\end{eqnarray}%
where $\{\gamma ^{\mu }\}_{\mu =1,..,4}$ are the Dirac matrices in the Weyl
basis, $b_{i}^{\ast }=b_{i}$ for $\alpha =1,2$ and $b_{i}^{\ast }=-b_{i}$
for $\alpha =3,4$ giving the representation of the internal symmetry of $%
Q_{\alpha }$ and $\left[ ,\right] $ the graded Lie bracket. Let us mention
that the supersymmetric generators $\{Q_{\alpha }\}$ are given in the
Majorana representations \cite{Wess}. Note that the Grassmann degrees of the
superfield components of $\Phi $ are\ given by $\mid \Phi _{M}^{i}\mid =\mid
\Phi _{M}^{\mu }\mid =m,$ $\mid \Phi _{M}^{\alpha }\mid =m+1$ $(\func{mod}2$
), where $m=\mid z^{M}\mid (m=0$ for $M$=$\mu $ and $m=1$ for $M=\alpha ),$
since $\Phi $ is an even 1-superform.

However, we assign to the anticommuting coordinates $\theta ^{1}$ and $%
\theta ^{2}$ the ghost numbers $(-1)$and $(+1)$ respectively, and ghost
number zero for an even quantity:\ either a coordinate, a superform or a
generator. These rules permit us to determine the ghost numbers of the
superfields $(\Phi _{\mu }^{i},$ $\Phi _{\mu }^{\nu },$ $\Phi _{\mu
}^{\alpha },$ $\Phi _{1}^{i},$ $\Phi _{2}^{i},$ $\Phi _{1}^{\nu },$ $\Phi
_{2}^{\nu },$ $\Phi _{1}^{\alpha },$ $\Phi _{2}^{\alpha })$ which are given
by $(0,$ $0,$ $0,$ $+1,$ $-1,$ $+1,$ $-1,$ $+1,$ $-1).$

Upon expressing the supercurvature $\Omega $ as

\begin{equation}
\Omega =\frac{1}{2}dz^{N}\wedge dz^{M}\Omega _{MN}=\frac{1}{2}dz^{N}\wedge
dz^{M}(\Omega _{MN}^{i}I_{i}+\Omega _{MN}^{\mu }P_{\mu }+\Omega
_{MN}^{\alpha }Q_{\alpha }),  \label{eq4}
\end{equation}%
we find\ from the structure equation 
\begin{subequations}
\begin{eqnarray}
\Omega _{\mu \nu } &=&\partial _{\mu }\Phi _{\nu }-\partial _{\nu }\Phi
_{\mu }+\left[ \Phi _{\mu },\Phi _{\nu }\right] ,  \label{Eq5a} \\
\Omega _{\mu \alpha } &=&\partial _{\mu }\Phi _{\alpha }-\partial _{\alpha
}\Phi _{\mu }+\left[ \Phi _{\mu },\Phi _{\alpha }\right] , \\
\Omega _{\alpha \beta } &=&\partial _{\alpha }\Phi _{\beta }+\partial
_{\beta }\Phi _{\alpha }+\left[ \Phi _{\alpha },\Phi _{\beta }\right] .
\end{eqnarray}%
Similarly, the Bianchi identity becomes

\end{subequations}
\begin{subequations}
\label{bianchi_4a}
\begin{gather}
D_{\mu }\Omega _{\nu \kappa }+D_{\kappa }\Omega _{\mu \nu }+D_{\nu }\Omega
_{\kappa \mu }=0,  \label{Eq bian} \\
D_{\alpha }\Omega _{\mu \nu }-D_{\nu }\Omega _{\mu \alpha }+D_{\mu }\Omega
_{\nu \alpha }=0, \\
D_{\alpha }\Omega _{\beta \gamma }+D_{\beta }\Omega _{\alpha \gamma
}+D_{\gamma }\Omega _{\alpha \beta }=0, \\
D_{\mu }\Omega _{\alpha \beta }-D_{\alpha }\Omega _{\mu \beta }-D_{\beta
}\Omega _{\mu \alpha }=0,
\end{gather}%
where $D_{M}=\partial _{M}+\left[ \Phi _{M},.\right] $ is the $M$ covariant
superderivative. Now, we shall search for the constraints to the
supercurvature $\Omega $ in which the consistency with the Bianchi
identities $(7)$ is ensured. This requirement insures then the off-shell
nilpotency of the BRST and anti-BRST algebra. The full set\ of
supercurvature constraints turns out to be given by

\end{subequations}
\begin{equation}
\Omega _{\mu \alpha }=0,\text{ \ \ \ \ \ \ \ }\Omega _{\alpha \beta }=0.
\label{contraintes}
\end{equation}%
It is easy to check the consistency of this set of supercurvature
constraints through an analysis of the Bianchi identities. Indeed, we remark
that identities $(7c)$ and $(7d)$ are automatically satisfied because of the
constraints (\ref{contraintes}) while identities $(7a)$ and $(7b)$ yield a
further restriction on supercurvature $\Omega $

\begin{eqnarray}
\Omega _{\mu \nu \mid }^{i} &=&F_{\mu \nu }^{i}=\partial _{\mu }A_{\nu
}^{i}-\partial _{\nu }A_{\mu }^{i}+\left[ A_{\mu },A_{\nu }\right] ^{i}, 
\notag \\
\Omega _{\mu \nu }^{\kappa } &=&0,\text{ \ \ \ \ \ \ \ \ \ }\Omega _{\mu \nu
}^{\alpha }=0.  \label{CONTRAINT}
\end{eqnarray}%
At this point, let us mention that the consistency of the horizontability
conditions (\ref{contraintes}) and (\ref{CONTRAINT}) with the Bianchi
identities $(7),$ as we will see later, guarantees automatically the
off-shell nilpotency of the BRST and anti-BRST transformations on all the
fields belonging to $\mathcal{N}=1$ super Yang-Mills theory.

\section{Auxiliary fields}

In order\ to derive the off-shell BRST structure of$\ \mathcal{N}=1$ super
Yang-Mills theory using the above BRST superspace formalism, it is necessary
to give the geometrical interpretation of the fields occurring in the
quantization of such theory. Besides the gauge potential $\Phi _{\mu \mid
}^{i}=A_{\mu }^{i},$ there exists the superpartener $\Phi _{\alpha \mid
}^{i}=\lambda _{\alpha }^{i}$ of $A_{\mu }$ which is introduced via the
field redefinition $\Phi ^{\rho i}=-(\gamma ^{\mu })^{\rho \sigma }/4\left[
Q_{\sigma },\Phi _{\mu }^{i}\right] ,$ $\Phi _{\mu \mid }^{\nu }=0$
representing the fact\ that the gauge potential associated to the
translation will not exist in the super Yang-Mills theory and an auxiliary%
\footnote{%
By definition an auxiliary field does not describe an independent degree of
freedom; its equation of motion is algebraic.} real field $\Phi _{\mid
}^{i}=\Lambda ^{i}$ which is required for the construction of off-shell
nilpotent BRST and anti-BRST transformations and introduced via the
supersymmetric transformations of the superpartener of the gauge potential

\begin{equation}
\Phi ^{i}=\delta _{\sigma }^{\rho }/4\left[ Q_{\rho },\Phi ^{\sigma i}\right]
.  \label{auxiliary}
\end{equation}%
Furthermore, we introduce the following now: $\Phi _{1\mid }^{i}=c_{1}^{i}$
is the ghost for Yang-Mills symmetry, $\Phi _{2\mid }^{i}=\overline{c}%
_{2}^{i}$ is\ the antighost of $c_{1}^{i},$ $B^{i}=\partial _{1}\Phi _{2\mid
}^{i}$ is the associated auxiliary field, $\Phi _{1\mid }^{\alpha }=\chi
_{1}^{\alpha }$ is the supersymmetric ghost, $\Phi _{2\mid }^{\alpha }=%
\overline{\chi }_{2}^{\alpha },$ is\ the antighost of $\chi _{1}^{\alpha },$ 
$G^{\alpha }=\partial _{1}\Phi _{2\mid }^{\alpha }$ is the associated
auxiliary field$,$ $\Phi _{1\mid }^{\mu }=\xi _{1}^{\mu }$ is the
translation symmetry ghost, $\Phi _{2\mid }^{\mu }=\overline{\xi _{2}^{\mu }}%
,$ is\ the antighost of $\xi _{1}^{\mu }$ and $E^{\mu }=\partial _{1}\Phi
_{2\mid }^{\mu }$ is the associated auxiliary field$.$ Let us mention that
the symmetry ghosts antighosts $\chi _{\rho }^{\alpha }$ are commuting
fields while the others $c_{\rho }^{i}$ and $\xi _{\rho }^{\mu }$ are
anticommuting. Note that the symbol $``\mid "$\ indicates that the
superfield is evaluated at $\theta ^{\alpha }=0.$ We also realize the usual
identifications: $Q_{\alpha }(X_{\mid })=\partial _{\alpha }X_{\mid },$
where $X$ is any superfield $Q=Q_{1}$ and $(\overline{Q}=Q_{2})$ is the BRST
(anti-BRST) operator.

The action of the $\mathcal{N=}1$ supersymmetric generators $\{P_{\mu
},Q_{\alpha }\}_{\mu =1,...,4;\alpha =1,...,4}$ on these fields is given by

\begin{eqnarray*}
\left[ P_{\mu },X\right] &=&\partial _{\mu }X, \\
\left[ Q_{\alpha },A_{\mu }^{i}\right] &=&-(\gamma _{\mu })_{\alpha \beta
}\lambda ^{\beta i}, \\
\left[ Q_{\alpha },\lambda ^{\beta i}\right] &=&-\frac{1}{2}(\sigma ^{\mu
\nu })_{\alpha }^{\beta }F_{\mu \nu }^{i}+\delta _{\alpha }^{\beta }\Lambda
^{i},
\end{eqnarray*}

\begin{eqnarray}
\left[ Q_{\alpha },c_{\rho }^{i}\right] &=&2(\gamma ^{\mu })_{\alpha \beta
}\chi _{\rho }^{\beta }A_{\mu }^{i},  \notag \\
\left[ Q_{\alpha },\Lambda ^{i}\right] &=&\gamma ^{\mu }D_{\mu }\lambda
_{\alpha }^{i},  \notag \\
\left[ Q_{\alpha },F_{\mu \nu }^{i}\right] &=&(\gamma _{\mu })_{\alpha \beta
}(D_{\nu }\lambda )^{\beta i}-(\gamma _{\nu })_{\alpha \beta }(D_{\mu
}\lambda )^{\beta i},  \label{eqalgeb1}
\end{eqnarray}%
where $D_{\mu }=\partial _{\mu }+\left[ A_{\mu },.\right] $ and $X$ any
fields.

It is worthwhile to mention that we are interested in our present
investigation on the global supersymmetric transformations, so that the
parameters of the $\mathcal{N}=1$ supersymmetric and translation groups must
be space-time constant, i.e.

\begin{eqnarray}
\partial _{\mu }\chi _{\alpha }^{\rho } &=&0,  \notag \\
\partial _{\mu }\xi _{\alpha }^{\nu } &=&0.  \label{const1}
\end{eqnarray}

Using the above identifications with (\ref{const1}) and inserting the
constraints (\ref{contraintes}) and (\ref{CONTRAINT}) into the Eqs. $(6)$
and $(7b),$ we obtain

\begin{eqnarray}
\partial _{\alpha }\Phi _{\mu \mid }^{i} &=&(D_{\mu }c_{\alpha })^{i}-\xi
_{\alpha }^{\nu }\left[ P_{\nu },A_{\mu }^{i}\right] -\chi _{\alpha }^{\rho }%
\left[ Q_{\rho },A_{\mu }^{i}\right] ,  \notag \\
\partial _{\alpha }\Phi _{\beta \mid }^{i}+\partial _{\beta }\Phi _{\alpha
\mid }^{i} &=&-\left[ c_{\alpha },c_{\beta }\right] ^{i}-\xi _{\alpha }^{\nu
}\left[ P_{\nu },c_{\beta }^{i}\right] -\chi _{\alpha }^{\rho }\left[
Q_{\rho },c_{\beta }^{i}\right]  \notag \\
&&-\xi _{\beta }^{\nu }\left[ P_{\nu },c_{\alpha }^{i}\right] -\chi _{\beta
}^{\rho }\left[ Q_{\rho },c_{\alpha }^{i}\right] ,  \notag \\
\partial _{\alpha }\Phi _{\beta \mid }^{\gamma }+\partial _{\beta }\Phi
_{\alpha \mid }^{\gamma } &=&0,  \notag \\
\partial _{\alpha }\Phi _{\beta \mid }^{\nu }+\partial _{\beta }\Phi
_{\alpha \mid }^{\nu } &=&-2\chi _{\alpha }^{\rho }(\gamma ^{\nu })_{\rho
\sigma }\chi _{\beta }^{\sigma },  \notag \\
\partial _{\alpha }\Phi _{\mid }^{\kappa i} &=&-\left[ \lambda ^{\kappa
},c_{\alpha }\right] ^{i}-\xi _{\alpha }^{\nu }\left[ P_{\nu },\lambda
^{\kappa i}\right] -\chi _{\alpha }^{\rho }\left[ Q_{\rho },\lambda ^{\kappa
i}\right] +\chi _{\alpha }^{\kappa }\Lambda ^{i},  \notag \\
\partial _{\alpha }\Omega _{\mu \nu \mid }^{i} &=&-\left[ c_{\alpha ,}F_{\mu
\nu }^{i}\right] -\xi _{\alpha }^{\tau }\left[ P_{\tau },F_{\mu \nu }^{i}%
\right] -\chi _{\alpha }^{\rho }\left[ Q_{\rho },F_{\mu \nu }^{i}\right] , 
\notag \\
\chi _{\alpha }^{\rho }(\partial _{\beta }\varphi _{\mid }^{i})+\chi _{\beta
}^{\rho }(\partial _{\alpha }\varphi _{\mid }^{i}) &=&-\chi _{\alpha }^{\rho
}\left( \left[ c_{\beta },\Lambda \right] ^{i}+\xi _{\beta }^{\nu }\left[
P_{\nu },\Lambda ^{i}\right] +\chi _{\beta }^{\rho }\left[ Q_{\rho },\Lambda
^{i}\right] \right)  \notag \\
&&-\chi _{\beta }^{\rho }\left( \left[ c_{\alpha },\Lambda ^{i}\right] +\xi
_{\alpha }^{\nu }\left[ P_{\nu },\Lambda ^{i}\right] +\chi _{\alpha }^{\rho }%
\left[ Q_{\rho },\Lambda ^{i}\right] \right) ,  \notag \\
\partial _{\alpha }\Phi _{\mid }^{\rho i} &=&(\gamma ^{\mu })^{\rho \sigma
}/4\left[ Q_{\sigma },\partial _{\alpha }\Phi _{\mu \mid }^{i}\right] , 
\notag \\
\partial _{\alpha }\Phi ^{i} &=&-\delta _{\sigma }^{\rho }/4\left[ Q_{\rho
},\Phi _{\sigma }^{i}\right] .  \label{Inser}
\end{eqnarray}

Inserting Eq. (\ref{eqalgeb1}) into (\ref{Inser}), and evaluating these at $%
\theta ^{\alpha }=0$, we find the following BRST transformations

\begin{eqnarray}
QA_{\mu }^{i} &=&D_{\mu }c^{i}-\xi ^{\rho }\partial _{\rho }A_{\mu
}^{i}+\chi \gamma _{\mu }\lambda ^{i},  \notag \\
Q\lambda _{\alpha }^{i} &=&-f_{jk}^{i}c^{j}\lambda _{\alpha }^{k}-\xi ^{\mu
}\partial _{\mu }\lambda _{\alpha }^{i}+\frac{1}{2}(\chi \sigma ^{\mu \nu
})_{\alpha }F_{\mu \nu }^{i}+\chi _{\alpha }\Lambda ^{i},  \notag \\
Qc^{i} &=&-\frac{1}{2}f_{jk}^{i}c^{j}c^{k}-\xi ^{\mu }\partial _{\mu
}c^{i}+\chi \gamma ^{\mu }\overline{\chi }A_{\mu }^{i},  \notag \\
Q\xi ^{\rho } &=&-\chi \gamma ^{\rho }\overline{\chi },  \notag \\
Q\Lambda ^{i} &=&-f_{jk}^{i}c^{k}\Lambda ^{j}-\xi ^{\rho }\partial _{\rho
}\Lambda ^{i}-\chi \gamma ^{\mu }D_{\mu }\lambda ^{i},  \notag \\
QF_{\mu \nu }^{i} &=&-f_{jk}^{i}c^{k}F_{\mu \nu }^{j}-\xi ^{\rho }\partial
_{\rho }F_{\mu \nu }^{i}-\chi ^{\rho }\left\{ (\gamma _{\mu })_{\rho \sigma
}(D_{\nu }\lambda )^{\sigma i}-(\gamma _{\nu })_{\rho \sigma }(D_{\mu
}\lambda )^{\sigma i}\right\}  \notag \\
Q\chi ^{\alpha } &=&0,  \notag \\
Q\overline{c}^{i} &=&B^{i},  \notag \\
QB^{i} &=&0,  \notag \\
Q\overline{\xi }^{\mu } &=&E^{\mu },  \notag \\
QE^{\mu } &=&0,  \notag \\
Q\overline{\chi ^{\alpha }} &=&G^{\alpha },  \notag \\
QG^{\alpha } &=&0,  \label{brst -trans}
\end{eqnarray}%
and also the anti-BRST transformations, which can be derived from (\ref{brst
-trans})\ by the following mirror symmetry of the ghost numbers given by: $%
X\rightarrow X$ if $X=A_{\mu }^{i},$ $\lambda _{\alpha }^{i},$ $\Lambda
^{i}; $ $X\rightarrow \overline{X}$ if $X=Q,$ $c^{i},$ $B^{i},\xi ^{\mu
},E^{\mu },\chi ^{\alpha },G^{\alpha },$and $X=\overline{\overline{X}}$ where

\begin{eqnarray}
B^{i}+\overline{B^{i}} &=&f_{jk}^{i}c^{k}\overline{c}^{j}-\xi ^{\rho
}\partial _{\rho }\overline{c^{i}}-\overline{\xi }^{\rho }\partial _{\rho
}c^{i}-\chi \gamma ^{\mu }\overline{\chi }A_{\mu }^{i}-\overline{\chi }%
\gamma ^{\mu }\chi A_{\mu }^{i}, \\
E^{\mu }+\overline{E^{\mu }} &=&2\chi \gamma ^{\mu }\overline{\chi },  \notag
\\
G^{\alpha }+\overline{G^{\alpha }} &=&0.  \label{brst-trans2}
\end{eqnarray}

Let us note that the introduction of an auxiliary real field $\Lambda ^{i}$
besides the fields present in quantized $\mathcal{N}=1$ super Yang-Mills
theory in four-dimensions, guarantees\ automatically the off-shell
nilpotency of the$\left\{ Q,\overline{Q}\right\} $-algebra and make easier
then, as we will see in the next Section, the gauge-fixing process.

\section{Quantum action}

In the present section, we show how to construct in the context of our
procedure a BRST-invariant quantum action for $\mathcal{N}=1$ super
Yang-Mills theory as the lowest component of a quantum superaction. To this
purpose, let us recall that the gauge-fixing superaction similar to that
obtained in the case of Yang-Mills theories \cite{tah1,tah2} and
gauge-affine gravity \cite{meziane} is given by%
\begin{eqnarray}
S_{sgf} &=&\dint d^{4}xL_{sgf,}  \notag \\
L_{sgf} &=&(\partial _{1}\Phi _{2})(\partial ^{\mu }\Phi _{\mu })+(\partial
^{\mu }\Phi _{2})(\partial _{1}\Phi _{\mu })+(\partial _{1}\Phi
_{2})(\partial _{1}\Phi _{2}).  \label{fixing1}
\end{eqnarray}

We note first that in the case of Yang-Mills theory the superaction involve
a Lorentz gauge \cite{tah11} given by

\begin{equation}
\partial _{\mu }\Phi _{\mid }^{\mu }=0.  \label{gauge-fix1}
\end{equation}

In the case of super Yang-Mills theory we shall choose a supersymmetric
gauge-fixing which is the extension of the Lorentz gauge. This gauge fixing
can be obtained from (\ref{gauge-fix1}) by using the following substitution

\begin{equation}
\Phi _{\mu }\longrightarrow \widetilde{\Phi _{\mu }}=\Phi _{\mu }+\left[
\partial _{\mu }\Phi ^{\alpha },Q_{\alpha }\right] .  \label{gauge-fix2}
\end{equation}

Now, it is easy to see that the gauge-fixing superaction (\ref{fixing1}) can
be put in the following form

\begin{equation}
S_{sgf}=\dint d^{4}x(\partial _{1}\Phi _{2})(\partial ^{\mu }\widetilde{\Phi
_{\mu }})+(\partial ^{\mu }\Phi _{2})(\partial _{1}\widetilde{\Phi _{\mu }}).
\label{gauge-fix3}
\end{equation}

To determine the gauge-fixing action $S_{gf}$ as the lowest component of the
gauge-fixing superaction $S_{gf}=S_{sgf\mid }$, we impose the following rules%
\begin{eqnarray}
Tr(I^{m}I_{n}) &=&\delta _{n}^{m}  \notag \\
Tr(\left[ Q_{\alpha },Q_{\beta }\right] ) &=&2\gamma _{\alpha \beta }^{\mu }%
\mathcal{\partial }_{\mu }  \notag \\
Tr(P^{2}) &=&0.  \label{eqfix5}
\end{eqnarray}

These rules permit us to compute the trace of each term in (\ref{gauge-fix3}%
) . Indeed, from (\ref{eqfix5}) it is easy to put the gauge-fixing action $%
S_{^{gf}}$ in the form

\begin{eqnarray}
S_{gf} &=&S_{sgf\mid }=\dint d^{4}x(B\partial ^{\mu }A_{\mu }+2b_{j}^{\ast
}G(\gamma ^{\mu }\mathcal{\partial }_{\mu }\square \lambda ^{j})  \notag \\
&&+(\partial ^{\mu }\overline{c})\left\{ D_{\mu }c+\xi ^{\nu }\partial _{\nu
}A_{\mu }+\chi \gamma _{\mu }\lambda \right\}  \notag \\
&&-2b_{j}^{\ast }(\partial ^{\mu }\overline{\chi })\gamma ^{\nu }\mathcal{%
\partial }_{\nu }\partial _{\mu }\left\{ f_{ik}^{j}\lambda ^{i}c^{k}+\xi
^{\tau }\partial _{\tau }\lambda ^{j}+\frac{1}{2}\chi \sigma ^{\tau \upsilon
}F_{\tau \upsilon }^{i}+\chi D^{j}\right\} ).  \label{gaugefixac}
\end{eqnarray}

On the other hand, the presence of the extrafields breaks the invariance of
the classical action (\ref{Lagrangian}). In fact, the only terms which may
contribute to the $Q$-variation of the classical action $S_{0}$ are those
containing the extrafield $\Lambda ^{i}.$ This follows from the fact that
the BRST transformations up to terms $\Lambda ^{i}$ represent the $\mathcal{N%
}=1$ super Yang-Mills transformations expressed \`{a} la BRST. A simple
calculation with the help of the BRST transformations (\ref{brst -trans})
leads to%
\begin{equation}
QS_{0}=\Lambda ^{i}\gamma ^{\mu }D_{\mu }\lambda _{i}.  \label{action auxi}
\end{equation}%
Thus the classical action $S_{0}$ is note BRST-invariant, and in order to
find the BRST-invariant extension $S_{inv}$ of the classical action, we
shall add to $S_{0}$ a term $\widetilde{S_{0}}$ so that

\begin{equation}
Q(S_{0}+\widetilde{S_{0}})=0.  \label{Q(inva)}
\end{equation}%
To this end, we propose to write $\widetilde{S_{0},}$\ which define the
extension action for the auxiliary field $\Lambda ^{i}$ as follows

\begin{equation}
\widetilde{S_{0}}=-\frac{1}{2}\Lambda ^{i}\Lambda _{i}.  \label{aux actio22}
\end{equation}%
Then, it is quite easy to show that $Q(S_{0})=-Q(\widetilde{S_{0}})$ by a
direct calculation with the help\ of the transformations (\ref{brst -trans}).

Having found the BRST-invariant extension action $S_{inv}$ we now write the
full off-shell BRST- invariant quantum action $S_{q}$ by adding to the $Q$%
-invariant action $S_{inv}=S_{0}+\widetilde{S_{0}}$ the $Q$-gauge-fixing
action $S_{gf}$

\begin{equation}
S_{q}=S_{0}+\widetilde{S_{0}}+S_{gf}.  \label{quan act}
\end{equation}

Let us mention that the quantum action (\ref{quan act}) and the off-shell
BRST transformations (\ref{brst -trans}) obtained in the previous section
are equivalent to those proposed in \cite{Baulieu},\ where the
Batalin-Vilkovisky formalism has been considered to close the supersymmetric
algebras without relying on the unusual auxiliary fields.

It is worth nothing that the quantum action (\ref{quan act}) allows us to
see that the auxiliary field $\Lambda ^{i}$ does not propagate, as its
equation of motion is a constraint

\begin{equation}
\frac{\delta S_{q}}{\delta \Lambda ^{i}}=-\Lambda ^{i}+2b_{i}^{\ast }(%
\overline{\chi }\gamma ^{\mu }\mathcal{\partial }_{\mu }\square \chi )=0.
\label{eq motion}
\end{equation}

Thus the essential role of the nondynamical auxiliary field $\Lambda ^{i}$
is to close the BRST and anti-BRST algebra off-shell.

The elimination of the auxiliary field $\Lambda ^{i}$ by means of its
equation of motion (\ref{eq motion}) leads to the same gauge-fixed theory
with on-shell nilpotent BRST transformations obtained in the context of BV
formalism \cite{Baulieu} as well as in the framework of the superfibre
bundle approach \cite{tah1}.

Moreover, in our formalism we have also introduced an anti-BRST operator $%
\overline{Q}$ and it is important to realize that both the BRST symmetry and
anti-BRST symmetry can be taken into account on an equal footing. To this
end, we simply use the fact that there is a complete duality, with respect
to the mirror symmetry of the ghost number, between the $Q$- and $\overline{Q%
}$-transformations. So, the $\overline{Q}$-variation of the classical action 
$S_{0}$ is given by

\begin{equation*}
QS_{0}=\overline{\Lambda }^{i}\gamma ^{\mu }D_{\mu }\overline{\lambda }_{i}
\end{equation*}%
where $\overline{\Lambda }^{i}$ represents an auxiliary field in the context
of $\overline{Q}$-symmetry. Using however the $Q$-transformations of the
auxiliary field ( see Eqs. (\ref{brst -trans}) with the mirror symmetry), we
obtain that the $Q$-invariant action $S_{inv}$=$S_{0}+\widetilde{S_{0}}$ is
also $\overline{Q}$-invariant. Furthermore, the $Q$-gauge-fixing action can
be also written as in Yang-Mills theories in $\overline{Q}$-form. At this
point, we remark that the mirror symmetry allows to replace, in particular
the auxiliary field $\Lambda ^{i}$ in the context of $Q$-symmetry by $%
\overline{\Lambda }^{i}$. Therefore the full off-shell BRST-invariant
quantum action $S_{q}=S_{0}+\widetilde{S_{0}}+S_{gf}$ is also an off-shell
anti-BRST-invariant quantum action.

\section{\protect\bigskip Conclusion}

In this paper we have presented a BRST superspace approach in order to
perform the quantization of the four dimensional $\mathcal{N}=$1
supersymmetric Yang-Mills theory as model where the classical gauge algebra
is not closed. The construction is entirely based on the possibility of
introducing a set of auxiliary fields via the supersymmetric transformations
of the superpartener of the gauge potential associated to a super Yang-Mills
connection. The gauge fields and their associated ghost and antighost fields
occurring in\ quantized four dimensional $\mathcal{N}=1$ supersymmetric
Yang-Mills theory have been described through a super Yang-Mills connection,
whereas the extrafields coming from the supersymmetric transformations are
required to achieve the off-shell nilpotency of the BRST and anti-BRST
operators. The minimal set of extrafields is defined after having imposed
constraints on the supercurvature in which the consistency with the Bianchi
identities is guaranteed.

Furthermore, we have performed a direct construction of the BRST invariant
extension of the classical action for $\mathcal{N}=1,$ $4$ $D$
supersymmetric Yang-Mills theory in analogy with what it is realized in
simple supergravity \cite{tahiri22} and four-dimensional BF theories \cite%
{tahiri2}. The obtained quantum action allows us to see that the extrafields
enjoy the auxiliary freedom, i.e. their auxiliary fields do not propagate,
being vanishing on-shell. The elimination of these auxiliary fields\ using
the solution of their equations of motion permits us to recover the standard
quantum action with on-shell nilpotent BRST symmetry. The transformations of
this minimal set of auxiliary fields and the obtained BRST invariant action
agree with the standard results. By using the mirror symmetry between the
BRST and anti-BRST transformations, we can see that the BRST invariant
action is also anti-BRST invariant. Therefore the full quantum action is
BRST and anti-BRST invariant, since the gauge-fixing action can be written
as in the Yang-Mills case in BRST as well as anti-BRST exact form, due to
the off-shell nilpotency of the BRST-anti-BRST algebra.

Let us note, that the Batalin-Vilkovisky formalism can be used to obtain the
on-shell BRST invariant gauge fixed action for $\mathcal{N}=1$
supersymmetric Yang-Mills theory in four dimensions without requiring the
set of\ auxiliary fields but by extending the fields in the theory to
include the so-called antifields \cite{Baulieu}.

Finally, we should mention that the BRST superspace formalism presented in
this paper was successfully applied to several interesting theories: simple
supergravity \cite{tahiri22} and $4D$ non-Abelian BF theory where the
symmetry is reductible \cite{tahiri2}. Such formalism permit us to determine
the off-shell nilpotent BRST and anti-BRST algebra for gauge theories. In
particular, it gives another possibility leading to the minimal set of
auxiliary fields. Thus, it would be a very nice endeavour to use this basic
idea to study the structure of auxiliary fields in other gauge theories.
These are some of the issues that are under investigation at the moment.

\end{document}